%% file: main.tex
\documentclass[manuscript,screen]{acmart}
\usepackage{graphicx}
\usepackage{xcolor}
\usepackage{soul}

\AtBeginDocument{%
  \providecommand\BibTeX{{%
    \normalfont B\kern-0.5em{\scshape i\kern-0.25em b}\kern-0.8em\TeX}}}

\setcopyright{acmcopyright}
\copyrightyear{2024}
\acmYear{2024}
\acmDOI{10.1145/3613904.3642148}

\acmConference[CHI'24]{CHI Conference on Human Factors in Computing Systems}{May 11-16, 2024}{Hawaii, HI}
\acmBooktitle{CHI Conference on Human Factors in Computing Systems (CHI'24), May 11-16, 2024, 2024, Hawaii, HI, USA}
\acmPrice{15.00}
\acmISBN{978-1-4503-XXXX-X/18/06}

\begin{document}

\title{"I Got Flagged for Supposed Bullying, Even Though It Was in Response to Someone Harassing Me About My Disability.": A Study of Blind TikTokers' Content Moderation Experiences}

\author{Yao Lyu}
\orcid{0000-0003-3962-4868}
\affiliation{%
  \institution{Pennsylvania State University}
  \city{University Park}
  \state{Pennsylvania}
  \country{USA}
}
\email{yaolyu@psu.edu}

\author{Jie Cai}
\orcid{0000-0002-0582-555X}
\affiliation{%
  \institution{Pennsylvania State University}
  \city{University Park}
  \state{Pennsylvania}
  \country{USA}
}
\email{jpc6982@psu.edu}

\author{Anisa Callis}
\orcid{0009-0007-3322-3285}
\affiliation{%
  \institution{Pennsylvania State University}
  \city{University Park}
  \state{Pennsylvania}
  \country{USA}
}
\email{amc8276@psu.edu}

\author{Kelley Cotter}
\orcid{0000-0003-1243-0131}
\affiliation{%
  \institution{Pennsylvania State University}
  \city{University Park}
  \state{Pennsylvania}
  \country{USA}
}
\email{kcotter@psu.edu}

\author{John M. Carroll}
\orcid{0000-0001-5189-337X}
\affiliation{%
  \institution{Pennsylvania State University}
  \city{University Park}
  \state{Pennsylvania}
  \country{USA}
}
\email{jmcarroll@psu.edu}

\renewcommand{\shortauthors}{}


\input{Sections/0_Abstract}




\maketitle

\input{Sections/01_Intro}
\input{Sections/03_Related_Work}

\input{Sections/04_Methods}
\input{Sections/05_Findings}

\input{Sections/06_Discussion}
\input{Sections/08_Limitations}

\begin{acks}
Thanks for the reviewers' comments. 
\end{acks}

\bibliographystyle{ACM-Reference-Format}
\bibliography{BlindTok,modSLR,ref,BlindTok_1}

\end{document}

%% file: Sections/0_Abstract.tex
\begin{abstract}

The Human-Computer Interaction (HCI) community has consistently focused on the experiences of users moderated by social media platforms. Recently, scholars have noticed that moderation practices could perpetuate biases, resulting in the marginalization of user groups undergoing moderation. However, most studies have primarily addressed marginalization related to issues such as racism or sexism, with little attention given to the experiences of people with disabilities. In this paper, we present a study on the moderation experiences of blind users on TikTok, also known as "BlindToker," to address this gap. We conducted semi-structured interviews with 20 BlindTokers and used thematic analysis to analyze the data. Two main themes emerged: BlindTokers' situated content moderation experiences and their reactions to content moderation. We reported on the lack of accessibility on TikTok's platform, contributing to the moderation and marginalization of BlindTokers. Additionally, we discovered instances of harassment from trolls that prompted BlindTokers to respond with harsh language, triggering further moderation. We discussed these findings in the context of the literature on moderation, marginalization, and transformative justice, seeking solutions to address such issues.
\end{abstract}

%% file: Sections/01_Intro.tex
\section{Introduction}

Social media platforms provide users with the opportunity to enjoy, create, and share a diverse range of content. However, users occasionally post content that violates community guidelines, leading to content moderation, which involves the removal of content or suspension of user accounts by platforms. The field of Human-Computer Interaction (HCI) has consistently focused on understanding the experiences of users undergoing moderation on social media platforms. Research on these experiences encompasses various aspects, such as the reasons behind content moderation \cite{jhaver_did_2019}, the mechanisms employed for moderation \cite{chandrasekharan_quarantined_2022}, and user feedback regarding moderation \cite{jhaver_does_2019}. Recently, scholars have highlighted that while the stated goal of content moderation on certain platforms is to eliminate harmful content and foster a positive and inclusive online community, these moderation practices can inadvertently perpetuate biases, resulting in the marginalization of specific user groups \cite{ball-burack_differential_2021}. However, despite the extensive literature on marginalization resulting from content moderation, most studies tend to focus on marginalization related to factors such as racism \cite{ball-burack_differential_2021} or sexism \cite{paromita_shadowbanned_2022}. There has been limited attention given to the experiences of moderation and marginalization encountered by individuals with disabilities, such as visually impaired users. Visually impaired users constitute an active community on platforms like TikTok, with videos featuring hashtags like "BlindTok" (the name of the visually impaired user community on TikTok) and similar variants accumulating over 800 million views as of December 2023 \footnote{https://www.tiktok.com/tag/blindtok}. The significant reach of this community underscores the need for further exploration into their unique experiences on the platform.

\begin{figure}[htp]
    \centering
    \includegraphics[scale=0.3]{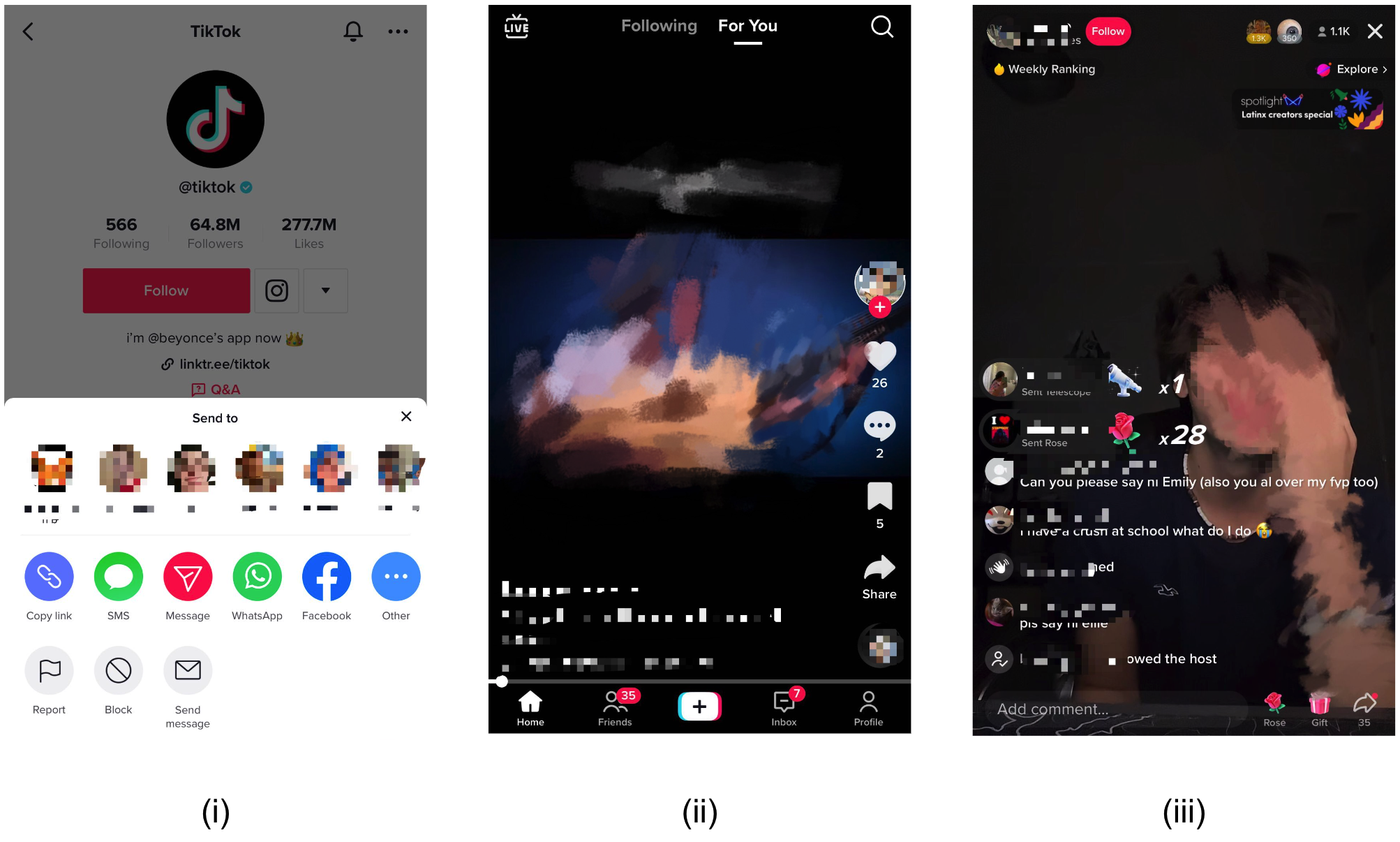}
    \caption{Three Mostly Used TikTok Interfaces (i) User Profile (ii) Video Interface and (iii) Live-streaming}
    \label{fig:1}
\end{figure}

This study is part of a long-term ethnographic project on the experiences of visually impaired users with TikTok. TikTok (\ref{fig:1}) is a popular short-video platform that encourages users to create and share content. Many platform features are designed to enhance the visual aspects of videos \cite{niu_building_2023,bartolome_literature_2023}. For instance, when checking a user's information, the icons are small (\ref{fig:1}, i); the like, comment, save, and share buttons are transparent and located on the left or bottom of the screen (\ref{fig:1}, ii); the comments on live streams float on the screen, and the font is small (\ref{fig:1}, iii). We utilized semi-structured interviews as the data collection method to understand how BlindTokers socialize, find entertainment, and learn on TikTok. During the initial interviews, BlindTokers spontaneously raised their experiences of being moderated by TikTok, which encouraged us to delve deeper into this topic in subsequent interviews. Prior research (\cite{devito_how_2022}) has examined the content moderation system on TikTok. In comparison to other platforms, TikTok employs a wide variety of algorithm-driven methods to manage content visibility, such as the "ForYouPage" (FYP) (\cite{simpson_for_2021}). The system is complex and often opaque to users, frequently resulting in negative emotions, such as feeling targeted or discriminated against, especially among minority groups (\cite{harris_honestly_2023}). In this study, we contribute to the exploration of users' experiences with the content moderation system on TikTok, using blind users as an example. We have specified two research questions to investigate these experiences (we have also designed interview questions based on these research questions; further details are presented in the methods section):

\begin{itemize}
\item RQ1: What are the moderation experiences of BlindTokers?
\item RQ2: What are the reactions of BlindTokers to the moderation experiences?
\end{itemize}


Ultimately, 20 participants shared their content moderation experiences with us. We employed thematic analysis \cite{Braun2006b} to analyze the data, leading to the identification of two themes related to our research questions: BlindTokers' contextual content moderation experiences and their responses to content moderation. We observed that accessibility issues played a pivotal role in triggering content moderation. The lack of accessible content creation tools made BlindTokers more susceptible to triggering the moderation system compared to sighted individuals, as exemplified by instances of posting content that included visual information violating copyrights. Furthermore, we identified other contributing factors to content moderation, including conflicts between BlindTokers and trolls. For instance, some BlindTokers resorted to using harsh language in response to trolls, consequently triggering the moderation system. We discussed these findings in the context of existing literature on content moderation and its implications for marginalization. Additionally, we applied a transformative justice framework \cite{mingus_transformative_2022,nocella_ii_overview_2011,gready_transformative_2010,daly_transformative_2002} to interpret situations where BlindTokers' content moderation experiences were provoked by troll harassment. In light of these insights, we propose design implications aimed at alleviating the aforementioned issues.


Our contributions to the literature encompass two primary aspects: 1) We identify diverse contexts in which BlindTokers encounter content moderation and respond to these experiences on TikTok, highlighting the shortcomings in TikTok's platform governance stemming from accessibility issues. 2) We document instances in which BlindTokers' content moderation was prompted by harassment from trolls. By examining these cases through the lens of transformative justice, we bring attention to underlying fairness issues at the societal/community level.

%% file: Sections/03_Related_Work.tex
\section{Related Work}
In this section, we review the literature concerning the experiences of blind individuals on social media, with a specific focus on accessibility design considerations previously examined in research. Additionally, we explore studies that delve into content moderation, particularly those practices that have led to the marginalization of minority groups. Lastly, we draw upon existing research on transformative justice within platform governance as a framework to analyze the interplay between content moderation and marginalization.

\subsection{Blind People and Social Media Accessibility}

Our study is situated within the context of prior research that delves into the experiences of individuals with visual impairments as they engage with social media platforms. Blind users' interactions with technology, especially the accessibility aspect, have always been an important topic for HCI researchers \cite{gleason_making_2019, macleod_understanding_2017, natalie_uncovering_2021}. Blind users, according to research reports, frequently grapple with the challenges posed by inaccessible visual content displayed on technology devices \cite{jaramillo-alcazar_mobile_2017, metatla_inclusive_2018}. Designers have responded to these challenges by developing technical solutions to enhance ICT accessibility \cite{jandrey_image_2021,liu_what_2021,wu_visually_2014}. However, accessibility problems cannot always be solved by technical solutions. For instance, a noteworthy challenge for accessibility arises from screen readers struggling to identify unlabeled buttons, an aspect that is ignored by developers when designing interface buttons \cite{ross_examining_2018}. Additionally, researchers have observed that visually impaired individuals' engagement with images is context-dependent, and the provided textual descriptions do not consistently align with their requirements \cite{abdolrahmani_empirical_2016, jung_so_2022, stangl_person_2020}. These accessibility challenges indicate that accessibility is not always a technical problem but also a social issue.

There has been a growing emphasis on video-oriented platforms used by blind individuals \cite{seo_understanding_2018, seo_understanding_2021, lee_online_2021, rong_it_2022}. On these platforms, users actively engage in creating or consuming video content and participate in live-streaming activities on platforms. Compared to text and image, these video-sharing platforms introduce a novel dimension in the consumption of content \cite{jun_exploring_2021}. Researchers have conducted an investigation into the interactions of individuals with visual impairments with live-streaming services and discovered that participants encountered substantial difficulties when it came to accessing information on these video-oriented platforms, such as difficulties in reading comments on the screen \cite{jun_exploring_2021}. Blind users had to find various ways to surmount the challenges, such as using external screen readers or seeking assistance from sighted individuals. 

In the present study, we extend our inquiry to encompass the experiences of blind users within the domain of short video platforms. This study is similar to what Rong et al. \cite{rong_it_2022} reported on blind users of Douyin, also recognized as the Chinese counterpart of TikTok. The authors meticulously documented how blind users in China reacted to Douyin's content recommendation algorithms. This exploration mentioned the users' experiences of being suppressed by the algorithm so they could not reach to a wider audience. While this paper has predominantly addressed accessibility issues within the broader scope of content creation, our unique focus is directed toward the specific content moderation encounters of BlindTokers. Our objective is to delve into the intricate interplay between accessibility challenges and content moderation systems, thereby offering valuable implications for future design considerations.

\subsection{Content Moderation and Marginalization}
Social media platforms are responsible for removing inappropriate content (e.g., hate speech, violence, harassment, etc.) and maintaining user engagement in a civil manner. Such content can cause various harms to those who experience it \cite{scott_trauma-informed_2023,scheuerman_framework_2021, schulenberg_creepy_2023} and to those who deal with it \cite{steiger_psychological_2021, wohn_volunteer_2019,dosono_moderation_2019}. Content moderation is defined as 'the practice of screening user-generated content (UGC) posted to Internet sites, social media platforms, and other online outlets, in order to determine the appropriateness of the content for a given site, locality, or jurisdiction' \cite{roberts_content_2017}. Platforms might view content moderation as a unique commodity and value proposition that they aim to deliver to the public \cite{riedl_book_2019}, employing both human moderators and algorithms to collaboratively work on filtering, reviewing, and removing content and accounts.

Many commercial platforms apply a centralized moderation mechanism, wherein the platforms control everything, such as defining rules and guidelines for their employees or training algorithms to reinforce the platform's values and vision \cite{gillespie_custodians_2018}. Algorithmic tools play a vital role in content moderation as they allow platforms to identify and react to harmful content at a large scale, such as de-recommending harmful videos on YouTube \cite{buntain_youtube_2021} and de-platforming toxic user groups on Reddit \cite{horta_ribeiro_platform_2021}. These tools can automatically detect and remove or flag content that violates community guidelines or policies for human moderators to evaluate. In this way, centralized moderation seeks to apply the same rules to all users universally.

However, recent work has shown that the 'one-size-fits-all' approach is not sustainable as communities with unique needs and interests grow and diversify \cite{schulenberg_towards_2023}. Many platforms empower their community members to moderate their own communities, such as volunteer moderators in subreddits on Reddit \cite{dosono_decolonizing_2020}, streamers' live streaming channels on Twitch \cite{cai_coordination_2022, cai_understanding_2023}, and administrators' server-based communities on Discord \cite{jiang_moderation_2019}. Platforms might provide some mechanisms to support user agency in the centralized moderation scheme, such as flagging content \cite{kou_flag_2021, crawford_what_2016}, reporting violators \cite{thomas_its_2022}, and appealing moderation \cite{banchik_disappearing_2021} if they are reported by others or moderated by the platform, even offering some personalized tools, such as word filters in communication settings \cite{uttarapong_harassment_2021}.

Algorithms developed by platforms often overly moderate content, especially sexual content such as nudity and body appearance \cite{are_emotional_2023}. A lack of transparency in the algorithm can lead to confusion among users who may not understand why their content was removed while other offensive content was left untouched \cite{ma_i_2022, register_attached_2023}. A thread of scholars highlights the importance of algorithm transparency to maintain users; vaguely communicated and/or weakly justified moderation decisions often leave users feeling unfairly treated \cite{jhaver_does_2019, ma_transparency_2023, wright_recast_2021}. To avoid content moderation, users develop various strategies. Some users might directly drop out of communities, or alter their content to circumvent content moderation when they feel irritated by unjust moderation \cite{chancellor_thyghgapp_2016}. Others involve coping strategies and labor to mitigate content moderation and preserve content, such as involving algorithmic labor and sharing their algorithm knowledge with the communities \cite{ma_how_2021, ma_i_2022}. Some employ folk theories to navigate systems to improve the visibility of their content \cite{devito_how_2022,devito_adaptive_2021}. Some preserve content in advance and appeal to the platform \cite{banchik_disappearing_2021}. Additionally, marginalized communities are more susceptible to targeted harassment and abuse, and content moderation processes may disproportionately affect them \cite{cai_hate_2023, han_hate_2023}. Algorithmic moderation is often biased to the detriment of marginalized groups \cite{blackwell_classification_2017,goyal_you_2022} and disproportionately produces false positives for marginalized groups \cite{haimson_disproportionate_2021}, such as YouTube's demonetization of videos featuring LGBTQ+ issues \cite{kingsley_give_2022}. Such biased moderation can limit their autonomy to express themselves and participate fully in the public sphere, perpetuating inequality and discrimination.

Recently, the HCI community has been paying growing attention to the interplay between marginalization and moderation specifically on TikTok. TikTok has been pushing content creators to keep producing content to maintain their visibility \cite{simpson_rethinking_2023}. However, extensive work has shown that TikTok's designs are problematic and could lead to the marginalization of certain populations \cite{lyu_because_2024,simpson_hey_2023}. For instance, Simpson et al. \cite{simpson_hey_2023} reported that creators with ADHD struggled to maintain their productivity due to the inaccessible designs of TikTok. Furthermore, marginalization could also be exemplified by content moderation/curation. Another two studies by Simpson and collaborators revealed that TikTok's algorithm-driven system could marginalize LGBTQ+ people by excluding content related to LGBTQ+ identity or pushing misleading content on it \cite{simpson_for_2021,simpson_how_2022}. Besides, Harris et al. \cite{harris_honestly_2023} presented a study on Black TikTok creators' content moderation experiences. The study reported that Black creators often got moderated, and the reasons for moderation revealed racial bias against Black people. For instance, creators who used African-American Vernacular English (AAVE) would be identified as hate speech.

In this study, we focus on the content moderation experiences of people with disabilities; specifically, we select BlindTokers as an example and explore how they are marginalized by TikTok's platform governance. We might expect to see similar issues with content moderation as have been documented for other marginalized groups, but likely not in the same ways. Investigating BlindTokers' experiences with content moderation can help reveal and correct any problems that would suppress their participation and representation on the platform.

\subsection{Transformative Justice in HCI}

We also use justice theories to examine BlindTokers' content moderation experiences. Justice is one of the fundamental desires of individuals in modern society; people aspire to be treated equally and without biases. In HCI, scholars have conducted numerous studies to examine justice issues occurring during human interactions with ICTs. Some are interested in justice issues engendered by technical designs, especially in algorithmic systems \cite{salac_scaffolding_2023}. Additionally, some scholars pay attention to exploring the effects of justice \cite{bosley_healing_2022}. Multiple studies \cite{xiao_addressing_2023,xiao_sensemaking_2022} have investigated how to address harm in online communities from a \textit{restorative justice} perspective. Restorative justice focuses on healing victims by addressing the harms with collective work that involves victims, offenders, facilitators, and the overall community \cite{xiao_sensemaking_2022}. This work includes understanding victims' needs, making violators understand their harmful behaviors, and encouraging community support for this process \cite{xiao_addressing_2023,xiao_sensemaking_2022}.

Additionally, there is a line of research specifically on justice issues in content moderation contexts \cite{xiao_addressing_2023,wang_how_2022,ball-burack_differential_2021}. As mentioned in the last subsection, many platforms use content moderation to regulate the information circulated on the platform to foster an inclusive atmosphere and reduce harmful content. However, some moderation practices can lead to community members feeling unfairly treated or even discriminated against. For example, Ball-Burack et al. \cite{ball-burack_differential_2021} presented a study on the \textit{racial dialect bias} of content moderation systems, revealing that tweets by Black creators, mostly using the dialect 'AAVE', were more likely to be identified as harmful content by automated moderation mechanisms, similar to the findings of Harris et al. \cite{harris_honestly_2023} on TikTok.

In this paper, we also draw from a particular justice framework, transformative justice, to discuss BlindTokers' moderation experiences. Transformative justice is a framework that analyzes justice from a holistic perspective \cite{mingus_transformative_2022,nocella_ii_overview_2011,gready_transformative_2010,daly_transformative_2002}. It posits that offenders' harmful behavior is a consequence of the injustice of the overall system; systematic oppression or violence against the offenders leads to harmful behaviors. Therefore, transformative justice focuses on reconciliation rather than punishment. The operationalization of transformative justice involves three main approaches \cite{mingus_transformative_2022,nocella_ii_overview_2011,gready_transformative_2010,daly_transformative_2002}: 1) investigating the stakeholders of violations; 2) identifying the broader social context problems that lead to violations; 3) addressing the violations by resolving the social problems. Transformative justice frameworks have been used in many HCI studies to combat societal-level justice issues \cite{dickinson_amplifying_2021,sultana_shishushurokkha_2022,musgrave_experiences_2022,erete_applying_2021,sultana_unmochon_2021}, including issues like child sexual abuse \cite{sultana_shishushurokkha_2022}, domestic violence \cite{rabaan_survivor-centered_2023}, and racism in educational contexts \cite{erete_applying_2021}.

In this study, we use transformative justice as a framework to understand BlindTokers' experiences of content moderation. Instead of focusing solely on moderation, we situate BlindTokers' experiences in a broader context and examine the underlying societal-level reasons for content moderation. We aim to provide design implications that contribute to not only a fairer moderation system but also a more inclusive social platform for diverse user groups. More details are elaborated in the discussion section.

%% file: Sections/04_Methods.tex
\section{Methods}
In the methods section, we introduce the background of the study. This research is a sub-study of a long-term ethnography study focused on the experiences of blind TikTok users. We provide detailed information about data collection, participant recruitment, and data analysis procedures. Additionally, we acknowledge the potential biases in our study and present our approaches to mitigate these biases.

\subsection{Study Background}

This study is part of a long-term ethnographic project on visually impaired users' experiences with TikTok. The project aimed at understanding how people with visual impairments use TikTok in various contexts, including socializing, entertainment, learning, etc. The project consisted of three primary parts:

1) \textbf{Reaching out to the targeted population.} To be more effective in sampling, we used TikTok as the only platform for participant recruitment, ensuring that all participants were active users. We created an account with the first author's real identity information; then we used keywords such as "blind," "visual impairment," and other similar phrases to search for and follow users who publicly disclosed their visual impairment in their user ID, profile, or post content. Due to TikTok's privacy policy, we could not send private messages to users who did not follow us. Therefore, we conducted a series of ice-breaking activities. We visited potential participants' videos and live-streams, leaving supportive comments like "That's great" or "Thanks for sharing!" Similar to prior work that used TikTok videos for recruitment \cite{devito_how_2022,simpson_for_2021}, we also posted videos presenting information about our project so people could better know us. After these activities, we created a reliable and trustworthy atmosphere for communication, and interested BlindTokers followed our account. Then we were able to send more detailed information through private messages. The participant recruitment part of our project involved immersing ourselves in BlindTokers' activities, negotiating with TikTok's privacy policies, and being reflexive during the entire process. It also adds to the growing discussion on methodologies for reaching out to and understanding marginalized populations on TikTok (e.g., queer methods in \cite{duguay_tiktoks_2023}).

2) \textbf{Data collection.} We used semi-structured interviews as the data collection method. Initially, we designed an interview protocol covering several basic topics related to how visually impaired people use TikTok. These topics included \textit{content creation and consumption} (e.g., "What content do you like most on TikTok?" "What kind of videos do you post on TikTok?"), \textit{social interactions} (e.g., "What kind of friends have you made on TikTok?" "How do you interact with your friends on TikTok?"), and \textit{accessibility issues} (e.g., "For visually impaired users, what are the accessibility issues in terms of consuming or creating content on TikTok?").

As we interviewed more participants, we found the protocol not comprehensive enough. So, we added more sections on specific aspects. For instance, multiple participants reported their experiences with content moderation on TikTok. We identified this as a salient pattern of usage among visually impaired users. Therefore, we added questions about \textit{content moderation experiences} (e.g., "What kind of moderation have you experienced?"), \textit{reasons for content moderation} (e.g., "From your perspective, why was your content moderated?"), \textit{consequences of content moderation} (e.g., "What happened after your content was moderated?"), and \textit{reactions to content moderation} (e.g., "How did you react to the moderation?"). During the interviews, we also asked about participants' perceptions of TikTok in these experiences, such as their views on the \textit{moderation system}, \textit{customer service}, and \textit{technical designs of TikTok}.

We spent a year in data collection and interviewed sixty visually impaired TikTokers. Among them, 20 reported experiences related to content moderation. We provided this information in Table \ref{table:demo} and labeled the participants as "C-number" ("C" stands for "creator"). These participants' ages ranged from 19 to 62 (avg=33). Regarding self-reported gender, 14 were female, 3 were male, and 3 were non-binary (Enby). We asked about participants' gender information without predefined options and reported it exactly as disclosed in the interviews. Educational backgrounds varied: five had high school diplomas, seven had associate degrees, one had a GED, four had bachelor's degrees, and three had master's degrees. In terms of visual impairments, three had low vision, fifteen were legally blind, and two were totally blind. All interviews were conducted in English, recorded as audio files, and then transcribed into text documents.

3) \textbf{Data analysis.} Considering the importance of the topic in HCI (content moderation in accessibility contexts), we decided to conduct a specific study based on the data from the 20 participants who reported content moderation experiences. We used thematic analysis \cite{Braun2006b} to analyze the data. In the first coding round, we focused on stories describing content moderation experiences, coding each part of the experience as "triggers of content moderation," "experiences of content moderation," and "feedback on content moderation." We then grouped all the codes and searched through them to identify salient patterns, noting that reasons for and reactions to content moderation were highly contextual. Consequently, we reported the content moderation experiences from a holistic perspective, covering not only the experiences but also the triggers and aftermath. The final report included two overarching themes: BlindTokers' situated content moderation experiences and their reactions to it.

\subsection{Ethical Considerations}
We also paid close attention to potential ethical issues and worked diligently to avoid them. We obtained approval from the Institutional Review Board (IRB) before starting participant outreach. During outreach, we provided our real names, email addresses from our institution, and the project's official website to ensure transparent and trustworthy communication. Since the study was initiated during the COVID-19 pandemic, all recruitment and interviews were conducted remotely to mitigate risks to participants. We interviewed all participants either via Zoom or phone calls. To address accessibility issues inherent in ICT-mediated communication, we primarily used verbal channels. For example, instead of using questionnaires to collect demographic information, we gathered this data through verbal communication before the interviews. Before each interview, we informed the interviewee of the research process, including data collection, storage, and potential privacy risks, and commenced the interview only after obtaining consent. After the interviews, we transcribed the audio files into text documents, ensuring all identifiable information, such as names or addresses, was removed during transcription.

Another critical ethical consideration involved the authors' backgrounds. All authors of this paper are sighted, and we acknowledge the potential bias due to the absence of blind authors in our team. In line with previous work \cite{xie_helping_2022}, we engaged a BlindToker, an undergrad student majoring in a relevant information technology field, as an advisor to our research team. We consulted this advisor throughout the process to better understand the study population, assist in participant recruitment, review our study methods, and provide feedback on our findings. For instance, we read the findings manuscript to the advisor, who helped us identify and eliminate any inappropriate phrases that might inadvertently other the blind participants. The consultant volunteered for this role.

Despite these precautions, the study still faced potential risks. For instance, we encouraged participants to invite their BlindToker friends to join the study. Although we instructed them to extend invitations privately, many participants chose to do so publicly on TikTok, posting videos to disclose their participation and enhance the project's credibility. These public videos also helped the project reach a wider audience and attract more participants. We are grateful for their efforts, which significantly aided our research, but we acknowledge the potential compromise of their anonymity in our project.

\begin{table}[]
\caption{Demographic Information}
\begin{tabular}{ccccc}
\hline
\textit{\#} & \multicolumn{1}{c}{\textit{Age}} &\multicolumn{1}{c}{\textit{Gender}}  & \multicolumn{1}{c}{\textit{Education}} & \multicolumn{1}{c}{\textit{Impairment}}  \\ 
\hline
C01            & 24  & Female      & High School      & Legally Blind \\
C02            & 39  & Female      & Associate & Legally Blind \\
C03            & 32  & Female      & Associate & Legally Blind \\
C04            & 29  & Male      & High School      & Legally Blind \\
C05            & 43  & Male     & Master    & Legally Blind \\
C06            & 32  & Female      & Bachelor  & Low Vision    \\
C07            & 59  & Female      & High School      & Totally Blind \\
C08            & 30  & Female      & Associate & Low Vision    \\
C09            & 32  & Enby   & High School      & Legally Blind \\
C10            & 24  & Male      & Bachelor  & Legally Blind \\
C11            & 19  & Enby   & High School      & Legally Blind \\
C12            & 23  & Enby   & Bachelor  & Totally Blind \\
C13            & 25  & Female      & Associate & legally blind \\
C14            & 21  & Female      & Bachelor  & Low Vision    \\
C15            & 62  & Female      & Associate & legally blind \\
C16            & 34  & Female      & GED       & Legally blind \\
C17            & 23  & Female      & Associate & Legally Blind \\
C18            & 27  & Female      & Associate & Legally Blind \\
C19            & 44  & Female      & Master    & Legally Blind \\
C20            & 25  & Female      & Master    & Legally Blind \\
\hline       

\end{tabular}
\label{table:demo}
\end{table}

%% file: Sections/05_Findings.tex
\section{Findings}

In this section, we report on two overarching themes related to BlindTokers' experiences with content moderation. The first theme explores the social and technical contexts of content moderation as described by the participants. This theme will delve into how the participants navigate the digital environment of TikTok, including the platform's design and functionality, and how these factors intersect with their visual impairments. The second theme examines the participants' reactions to content moderation. This includes their emotional and practical responses to moderated content, highlighting the unique challenges they face as visually impaired users. We will also explore their interactions with other BlindTokers and sighted TikTok users, emphasizing the social dynamics at play. Both themes collectively shed light on the multifaceted challenges faced by BlindTokers. These include the complexities of dealing with visual impairments, navigating social interactions on the platform, and contending with TikTok's often inaccessible design. The insights from these themes contribute to a deeper understanding of the user experience for visually impaired individuals on social media platforms. Further details will be presented in the following subsections.

\subsection{BlindTokers' Contextual Content Moderation Experiences}

From the participants' responses, it became clear that content moderation on TikTok occurred in various contexts. The first type identified was technically-induced moderation. This form of moderation is driven by algorithms that TikTok employs to automatically detect and remove content violating the community guidelines. Participants recounted instances where they inadvertently triggered these auto-moderation systems, attributing these occurrences to a lack of accessibility considerations within TikTok's design. The second type was socially-induced moderation, which arose from the social interactions between BlindTokers and the TikTok platform, as well as with sighted trolls. In this context, BlindTokers perceived a bias in the platform's moderation practices, feeling that TikTok was intentionally shadow-banning content created by blind users. Furthermore, BlindTokers frequently faced ableist harassment and trolling from other users. When they responded to such trolling with explicit or harsh language, they often found themselves banned by the platform. This dynamic highlights the complex interplay between the technical and social aspects of content moderation and its impact on the BlindToker community.

\subsubsection{\textbf{Technically Induced Moderation}}

Like most platforms, TikTok has implemented a complex content moderation system driven by algorithms. While this automated moderation helps manage content at scale, it often fails to accurately interpret nuanced human interactions. This can lead to misclassification of benign or otherwise unproblematic content as harmful, and vice versa, as reported in previous work \cite{haimson_disproportionate_2021}. Our participants (N=12) expressed similar frustrations with this automatic moderation system. One notable issue was the occurrence of technically-induced false positives, arising because the platform did not adequately consider the lifestyles of blind people. Blind individuals generally rely less on visual information from their surroundings compared to sighted individuals. Consequently, when recording videos, they may pay less attention to the visual aspects. If the platform lacks accessible designs that cater to BlindTokers, these users might unintentionally violate guidelines. For instance, participant C09 commented: 

\begin{quote}\textit{I know I did have some violations on that account, because of me being visually impaired, I'm (my face) often on top of my screen. And I used to wear a strapless dress. So I would get flagged for minor safety because they think I'm an underage naked person.} [C09]  \end{quote}

TikTok features "talking head" style videos, where individuals sit in the center of the frame, with their heads taking up most of the screen. However, when C09 created videos in this style, TikTok failed to provide accessible reminders that her head was not properly positioned in the middle. This resulted in unintentional nudity in her videos and eventually led to the moderation of such content.

Another scenario of technically induced moderation involved users getting banned due to captions in their videos. On TikTok, users have the option to manually add captions to their videos. Moreover, many BlindTokers informed us that they intentionally added captions to their videos as a way to demonstrate support and consideration for people with hearing impairments. C10 remarked:

\begin{quote}\textit{A lot of them (my videos) are me just kind of sitting and talking to the camera. So there isn't too much of a visual aspect unless you want to see my face. And I include captions for the people who have difficulty hearing or can't hear.} [C10] \end{quote}

Captions in a video help viewers see what the video is about. Among many TikTok creators, adding captions is a way to make videos more accessible to people with hearing impairments. BlindTokers were concerned about the overall accessibility issues on TikTok, not just the accessibility for blind people. Therefore, by adding captions, C10 also showed considerations for people with hearing impairments. However, the caring work to improve the accessibility for others (adding captions) sometimes brought them trouble. C09 told us: 

\begin{quote}\textit{Yes, I'm very picky about my captions...They are pre-filtering your captions, like your captions say something that triggers their guidelines, you're gonna get an automatic violation.} [C09] \end{quote}

C09 cautioned us that using captions would risk being flagged as a "violation" due to the automatic algorithms used for content moderation. C08, who had experienced abuse from her father during her childhood, posted a video sharing her experiences with the audience. However, her content was moderated by the automatic system. She later determined that it was the caption that triggered the moderation:

\begin{quote}\textit{I quoted him in the video. When I went to post it, it was immediately banned for hate speech in a split second, so nobody had time to report it. (Then I realized) I had the caption on, the caption itself instantly had me reported for his speech, so I posted the same one with no captions...and nothing happened.} [C08] \end{quote}

In her story, C08's video was flagged by the system as a violation due to captions that included quotes from her dad's hate speech. As a result, she had to disable the captions to avoid moderation. However, C08 also noted that the absence of captions created accessibility issues for her friend with hearing impairments. She had to send messages to her friend to explain what was happening in the videos that lacked captions. 

\subsubsection{\textbf{Socially Induced Moderation}}

In addition, the participants (N=15) also reported socially induced moderation. Unlike technically induced moderation caused by overly simplified and automatic moderation algorithms, socially induced moderation arises from various social factors, such as the platform's marginalization of BlindTokers and the ableism exhibited by sighted TikTokers towards BlindTokers. While moderation in this context was still carried out by automatic systems, which were also relevant to the technical side, social factors like marginalization and ableism were the primary reasons triggering the moderation. In other words, in the previous section, the problem was that while the automated process did detect legitimate cases of violation, the system did not adapt itself to accommodate BlindTokers' lifestyles, thus causing issues.

In this section, however, the problems revolved around the platform itself and the inherent bias of some sighted TikTokers against BlindTokers' identity as blind individuals. The moderation practices in this context were not even accurate. The first scenario illustrates instances where BlindTokers were moderated due to the platform's oppression of marginalized populations. C13 shared her experiences with us:

\begin{quote}\textit{...when I'm trying to do dance or something, that's when I get more shadowbanned of them. But I notice when I (just) use my voice, I get tons of views. So I'm just like, should I just talk more instead of putting (my face)?... And I feel like. okay, it doesn't want to show people because I'm blind.} [C13] \end{quote}

TikTok has been reported deliberately shadowbanning content created by various marginalized populations like LGBTQ+ people \cite{paromita_shadowbanned_2022}. BlindTokers also showed that their content would automatically get moderated because of their blind identity. Sometimes showing items that were closely relevant to blind people's lives, like white canes, also led to moderation ([C11]).

In addition to the platform's oppression, another set of moderation experiences stemmed from entrapment by ableists. By "entrapment," we refer to situations where certain sighted TikTokers deliberately engaged in harassment against BlindTokers and baited them into responding with harsh language. When BlindTokers did so, the moderation system would be triggered. C02 described how she was subjected to harassment through comments from some sighted TikTokers:

\begin{quote}\textit{Somebody, they ask questions about toileting, "How do you know when you are finished, make sure it's cleaned?" And I just thought, "Well, you’re an idiot."} [C02] \end{quote}

BlindTokers' activities on TikTok encompassed a diverse range of presentations aimed at providing insights into their lives as blind individuals, with the goal of fostering a better understanding of the blind community among sighted viewers. They shared the intricacies of living with visual impairments and addressed questions from sighted individuals about blind people's lifestyles. However, some people posed questions that appeared to be inquisitive but were actually intended to mock BlindTokers. C09 shared a story with us:

\begin{quote}\textit{I recently had a video that I was trying to stitch. It was a joke video with some chick talking with their friends like, "I never make blind friends", my response is just "Oh, honey and you sit down next to me, we're gonna talk about XYZ, don't make fun". And I posted the video on my main account, and it got BANNED. It got flagged and removed.} [C09] \end{quote}

C09 encountered various situations where blind people were harassed by sighted people, including people claiming they would not make blind friends and people ridiculing blind people by mimicking how blind people used white canes. She did not tolerate such harassment and chose to fight back. While she did not specify what she said in the video, the result of being banned indicated that she used harsh language. Actually, many blind people did the same thing, like C08 said: "\textit{I tell them off in a very, not nice way}." This would trigger the moderation system or give the trolls evidence to report the BlindTokers. C09 also told us what happened after she argued with an ableist, who made a video mimicking blind people:

\begin{quote}\textit{I told her that this was incredibly wrong. Then she tried to call me on everything. She had another video like, this whole blind community coming after them, and there could be death threats and stuff.} [C09] \end{quote}

Instead of offering apologies and attempting to resolve the conflicts, ableists often played the victim card and accused BlindTokers of violating community guidelines. Unfortunately, in many cases, the trolls who initiated the conflicts faced no consequences, while BlindTokers who defended themselves by fighting back would be reported or even banned. C01 shared her experiences:

\begin{quote}\textit{It's always frustrating, because if you make a video saying something, then TikTok feels like you're out of line, or you're harassing, then they'll either ban you or they take your video down. But (trolls) they don't get banned or anything for their rude comments.} [C01] \end{quote}

It's important to highlight that these trolls were adept at bypassing the moderation, and their actions couldn't be defined as rule violations since they didn't technically break any rules. As a result, BlindTokers ended up being labeled as violators and were subsequently punished by the platform. The unequal treatment of blind and sighted individuals in these conflicts left C01 feeling frustrated. The platform only penalized BlindTokers while sparing the trolls, possibly because blind individuals used overtly harsh language that was easily detectable, whereas the trolls' expressions were implicit and difficult for the platform to identify. Some participants recognized that the entire process, starting from harassment to moderation, served as a form of entertainment for sighted ableists. As C18 put it: "\textit{If I'd like to comment back, it's just kind of entertaining them, feeding into them, like they're getting the attention that they want.}"

\subsection{BlindTokers' Reactions To Content Moderation}

This subsection highlights the reactions of BlindTokers after experiencing content moderation. Content moderation held significant implications in the online lives of BlindTokers. Being subjected to moderation meant reduced visibility to their audiences, which adversely affected their everyday social interactions. For instance, C19, who enjoyed hosting live-streams for her audience, was shadowbanned. Shadowbanning is a form of moderation where an individual is not completely banned or removed from the platform; they can still post videos and host live-streams, but the platform reduces the exposure of their content to the broader audience, including their followers. Consequently, after being shadowbanned, C19's followers did not receive notifications from the platform. Instead, C19 had to manually invite people so they could join the live-streams. However, this process also involved navigating various inaccessible designs within TikTok:

\begin{quote}\textit{So I would have to send out the invitations otherwise they wouldn't have access, that was very frustrating...you have to go through each individual. And when you're using a screen reader, it slows the process down a lot, because you have to be like "Who's this person" and it's reading each person one at a time. And when people can see it, they just click and they can just keep going, but it's a little harder for us with visual impairment.} [C19] \end{quote}

As previously mentioned in the introduction section, TikTok poses numerous potential accessibility challenges. BlindTokers had reported a multitude of accessibility issues, such as the tiny profile icons that were problematic for individuals with visual impairments to discern \cite{jun_exploring_2021}. C19's story highlighted the disparities in remedial practices between sighted and blind TikTokers after experiencing moderation. Moreover, blind individuals had to overcome additional accessibility hurdles with extra effort.

Such frustrating experiences alerted BlindTokers to the fact that content moderation could have significant consequences, prompting them to handle it carefully. Overall, participants reported three types of reactions after experiencing content moderation. They shared their experiences with the TikTok platform in the hope of receiving fair treatment, but in most cases, their interactions with the TikTok platform proved unsatisfactory. Another approach was to discuss content moderation experiences with individuals within their social networks. BlindTokers deliberated on these experiences with their friends or sought assistance from fellow TikTokers, both sighted and blind. Lastly, some BlindTokers reported how they handled these issues on their own.

\subsubsection{\textbf{Handling with the Platform}}

When addressing content moderation, most participants (N=15) chose to contact TikTok directly to seek explanations for moderation decisions and to rectify misunderstandings that led to unwarranted actions. While some reported success in having their videos reinstated ([C05]), a significant number expressed dissatisfaction with the platform, mainly due to communication problems. Participants highlighted difficulties in accessing specific information about moderation, with C16, for example, finding it challenging to obtain precise details from the platform's notices.

\begin{quote}\textit{It alerted me... the one about harassment and bullying. I just got a little bell notification on TikTok and I had to go in and look at it. But that required a lot of link-clicking and forward finding. And I don't know how intuitive that would have been with VoiceOver.} [C16] \end{quote}

As we mentioned before, TikTok had been blamed by many BlindTokers due to its tiny icons. In C16's case, the way the notifications were sent to the users made the experience even worse. After being alerted by the platform about moderation, BlindTokers also tried to make sense of the rules of moderation. They then turned to the community guidelines of the platform. C03 recalled how he read the particular document:

\begin{quote}\textit{I skimmed through them. However, it’s very long and redundant. I learned by myself about making my VoiceOver at a pretty fast speed when I had to do that. But there are some people who use their VoiceOver and run a very slow speed and (when reading) they're like “Hurry up”... }[C03]\end{quote}

According to C03, reading community guidelines was challenging for people with visual impairments due to the inaccessible design (lengthy and redundant text). In addition to grappling with understanding the platform's notices and guidelines, participants frequently reported their experiences of being ignored or treated unfairly by the platform. C16 expressed frustration with the platform's handling of these situations:

\begin{quote}\textit{Definitely just (auto-reply) email, they also don't really have a person that goes over the complaints. I wrote out my objections. And then I didn't really hear back from them like "We're not gonna reinstate your videos".} [C16] \end{quote}

The absence of feedback and the lack of human interaction were sources of frustration for the participants. Even in instances where participants did hear back from the platform, the outcomes often seemed unfair. C18 shared a story in which they engaged in an argument with a troll but ended up being moderated:

\begin{quote}\textit{...in one of my videos, I got flagged for supposed bullying, even though it was in response to someone harassing me about my disability. And I tried to appeal to that video and they said that, "Hey, it's still wrong." I can't get that video back.} [C18] \end{quote}

In C18's situation, reporting the issue to the platform proved to be futile, despite receiving a response from the platform. TikTok continued to evaluate C18's videos without taking into account the full context of the story. After numerous disappointing experiences, participants lost trust in the platform. C11 went even further to share a conspiracy theory about the platform:

\begin{quote} \textit{You just have to find workarounds and loopholes to get your content out there. Because if you contact them about it, then they flagged your page as one that may be violating those censorship policies. And so you're just putting yourself on the radar for no reason.} [C11] \end{quote}

C11's explanation sheds light on how and why BlindTokers developed folk theories based on their interactions, which echoes previous work's findings \cite{jhaver_did_2019}. These theories reflect the strained relationships between the platform and the moderated BlindTokers.

\subsubsection{\textbf{Handling with Social Network}}

In addition to their experiences with the platform, some participants (N=8) also shared how they navigated content moderation issues with individuals in their social networks. Following moderation by TikTok and feeling frustrated by the platform's response, some BlindTokers took videos detailing their experiences. These videos often criticized the platform for incorrect moderation, unreasonable feedback on their appeals, and the unfairness of the contexts that triggered the moderation, particularly with respect to blind individuals. They also shared techniques and strategies for dealing with and circumventing the platform's moderation systems. For example, C06, who aimed to raise awareness about the seriousness of ableist comments, shared techniques for bypassing the moderation systems:

\begin{quote}\textit{I've been really noticing how much bullying has been going on, not just on this app but on social media in general. And I wanted to say, "If people kill themselves because of a comment..." I just follow what other people said, like you can't say, "kill" or ”murder.“ So I said, "If people 'unlife' themselves..."} [C06] \end{quote}

As mentioned earlier, some participants inadvertently triggered the moderation system by using certain words, and regardless of the context, if a video contained harmful language, it would be subject to moderation. This led to a rather ironic situation for C06. She wanted to alert her audience about the consequences of harassment, including self-harm and moderation. However, due to the fear of being moderated, she couldn't use the word "kill" in her video. Inspired by other TikTokers, she learned how to rephrase her message to avoid moderation, a technique known as "algospeak \cite{klug_how_2023}," which has become prevalent among content creators as a way to circumvent algorithmic moderation. Despite the efforts of BlindTokers and their friends, including some who were influencers with a substantial following, the impact on the platform appeared to be minimal. In addition to circumventing moderation, some participants were advised by their friends to avoid conflicts that could potentially trigger moderation. C07, who had many blind TikTok influencer friends, shared her experience:

\begin{quote}\textit{Another stupid thing (comment), like "Jesus, why you come out (if you are blind)"...I was gonna say something really harsh, but my good friend (a BlindTok influencer) stopped me. She's like, people are gonna harass you, she knows what's gonna happen. And I really don't need to get into that kind of drama. So I didn't post anything back. }[C07] \end{quote}

Some BlindTokers who had more experience in dealing with conflicts involving ableists understood that confronting the ableists could escalate the situation and potentially lead to unexpected arguments and moderation. As a result, they often advised BlindTokers who were newly encountering such individuals to remain calm and avoid engaging in conflicts. In addition to receiving techniques and advice from friends, one participant shared her experience of seeking assistance from her sighted friends for a form of retaliation. C17 explained:

\begin{quote}{\textit{There's a group of people who fight bullies on TikTok... I screenshot their (trolls) profile and send it to somebody in the group...When I show the picture of these things, he knows exactly what's going on. They will send the picture to somebody else...so they can go after them.}} [C17]\end{quote}

C17 introduced a unique strategy: instead of responding to trolls directly herself (which could lead to conflicts and moderation), she sought assistance from a group of sighted volunteers who actively fought against ableism on TikTok. These volunteers would take action on behalf of the BlindToker by addressing the trolls. This strategy showcased that BlindTokers, along with other TikTokers, were willing to seek justice independently of the platform, which was supposed to handle such issues. The routinized process also indicated that sighted TikTokers had developed strategies for confronting others without falling victim to moderation. This underscored the accessibility challenges faced by TikTok in terms of supporting BlindTokers in their efforts to fight back.

\subsubsection{\textbf{Handling By Themselves}}

Many participants (N=10) also shared how they managed issues on their own. Some of them reflected on specific situations and then devised techniques to either minimize the impact of moderation or avoid further moderation. These self-developed techniques highlighted the participants' comprehension of TikTok's algorithmic features, moderation systems, and accessibility designs. One type of technique involved limiting the consequences of content moderation. Many participants informed us that content moderation could sometimes result in the removal of their accounts. As a precautionary measure, they prepared a backup account, as explained by C09:

\begin{quote}\textit{I had a backup, I made the announcement on my main account that I had one, and ask people to follow it with this one... My username isn't that different. And I make sure the hashtag "BlindTok". That's the biggest hope.} [C09] \end{quote}

C09 had many friends with disabilities on TikTok, and they regularly created videos and participated in each other's live streams. If her account were to be banned, C09 and her friends would lose a vital means of staying connected. To prevent this, C09 took the proactive step of preparing an alternative account. She also employed techniques like using similar usernames and relevant hashtags to make it easier for her friends to find her on the new account. Another type of technique involved avoiding future content moderation. For example, participants steered clear of using words or functions that might trigger the moderation algorithm, such as employing "algospeaks" in videos ([C06]) or turning off captions ([C09]). Given that many instances of moderation stemmed from conflicts with sighted TikTokers, BlindTokers often chose to avoid confrontation. Some participants shared their experiences of deleting rude comments or blocking individuals who left such comments, rather than engaging in arguments with them. C11 explained:

\begin{quote}{\textit{I block people as well. Like if you delete their comment, and then they notice that you delete the comment, they'll come back and continue to comment... (About blocking) I have to go through different things on the person’s page to figure out where was it...I actually have accidentally "liked" their videos a few times when I was trying to either unfollow them or block them, it is embarrassing.} [C01]}\end{quote}

As reported by C01, trolls didn't just leave rude comments and move on; they returned to check if their comments were still visible. This behavior underscored the deliberate and targeted nature of the trolls' actions against people with disabilities. However, in such cases, TikTok's functions that BlindTokers relied on, such as deleting comments and blocking trolls, often proved ineffective. TikTok required BlindTokers to navigate through cumbersome and inaccessible processes, and sometimes the outcome would further hurt their feelings. Moreover, some participants who encountered rude comments during live-streams also reported that deleting comments could sometimes trigger screen readers to read the offensive comment aloud. This inadvertently aided the trolls and made BlindTokers feel even worse ([C20]). This highlighted compatibility issues between TikTok and accessibility tools, which could worsen the experience of dealing with content moderation.

After such disappointing experiences, many participants chose to tolerate the harassment and live with it. C19 expressed this sentiment:

\begin{quote}\textit{I try to keep my videos as neutral as possible. I try not to stir the pot too much because if you offend the wrong person, they're gonna report your video and it's gonna get banned. So if people write something rude to me, I used to reply to those comments, but I don't anymore.} [C19] \end{quote}

C19 not only abandoned the fight against trolls but also altered her video style. She began to self-censor her content to avoid potentially offending anyone. These actions and thought processes illustrated BlindTokers' passive acceptance of a life where they often faced harassment from others.

%% file: Sections/06_Discussion.tex
\section{Discussion}

In the findings, BlindTokers encountered various instances of content moderation that made them feel marginalized by TikTok: 1) \textit{Being confused by inaccessible designs}. TikTok's functions supporting content creation were mostly inaccessible to blind users. The platform's lengthy community guidelines exacerbated this issue, creating an environment of constant confusion. 2) \textit{Being punished due to unintentional mistakes}. Navigating the confusing platform, compounded by inaccessibility, led BlindTokers to make unintentional mistakes. The inaccessible designs made them less aware of the visual aspects, resulting in violations. 3) \textit{Being ignored when protesting}. Facing moderation consequences, BlindTokers advocated for understanding, knowledge, and collaboration with the platform. However, feedback was limited. 4) \textit{Being forced to passively accept marginalization}. BlindTokers, losing trust in the platform, took matters into their own hands. They avoided conflicts, established alternative channels, and employed self-censoring techniques to evade moderation. This passive acceptance of marginalization hindered their freedom to create and share content on TikTok, further normalizing their marginalization on the platform.

TikTok has faced criticism for its disproportionate moderation practices against minority groups, such as LGBTQ+ \cite{paromita_shadowbanned_2022} and Black individuals \cite{ball-burack_differential_2021}, leading to increased marginalization. Our research contributes to this line of inquiry, highlighting the context of disability and accessibility. In other words, while some moderation experiences (such as copyright take-downs or accidental nudity \cite{harris_honestly_2023}) are common across different user groups, the implications of these experiences differ. For instance, the stories about captioning and moderation discussed in section 4.1.1 reveal two aspects distinct from other TikTokers' experiences. First, captioning represents a method of creating \textit{accessible} content on TikTok; second, individuals from \textit{marginalized groups} often share their stories on social media \cite{harris_honestly_2023}, and these stories might include sensitive words that trigger the moderation algorithm. Therefore, while other TikTokers may encounter similar experiences (moderation triggered by captions), the BlindTokers' cases specifically unveil TikTok's marginalization against one particular group: people with disabilities.

Furthermore, our findings are in line with the ongoing examination of algorithmic experiences among marginalized groups. Choi et al. \cite{choi_its_2022} reported that YouTubers with disabilities often experienced algorithmic suppression, suspecting that their content related to their marginalized identity (as people with disabilities) was excluded by the YouTube algorithm. Karizat et al. \cite{karizat_algorithmic_2021} investigated how users responded to such algorithmic suppression. Participants developed various folk theories and engaged in individual and collective actions to resist the suppression. Our study contributes to this body of research by presenting a more severe scenario, where users subject to moderation simply gave up resisting and resorted to self-censorship. This tendency may stem from the conflicts between BlindTokers and trolls, which further exacerbated their experiences of marginalization. We discuss this particular issue in later sections and provide design implications to mitigate these problems.

\subsection{BlindTokers' Moderation Experiences and Trolls' Harassment: A Transformative Justice Perspective}


While we have discussed BlindTokers' moderation experiences and TikTok's platform governance, this section focuses on moderation experiences resulting from trolls' harassment. In the previous section, we addressed moderation due to unintentional violations, mostly triggered by TikTok's inaccessible designs. However, we also observed moderation instances not stemming from technical design issues but from social interactions with trolls perpetuating ableism. As defined by disability justice activists, ableism is "\textit{a system that places value on people’s bodies and minds based on societally constructed ideas of normalcy, intelligence, excellence, and productivity.}" \cite{lewis_ableism_2020} Most participants frequently encountered harassment on TikTok. When individuals with limited understanding of visual impairments encountered BlindTokers' content, they often attacked them, accusing them of faking their disabilities. This reaction was rooted in the trolls' misaligned perceptions of blindness and a general lack of public awareness about visual impairments and assistive technologies. Consequently, BlindTokers who faced harassment from trolls often retaliated with harsh language, which in turn triggered content moderation.

In a recent literature review on content moderation, Jiang et al. \cite{jiang_trade-off-centered_2023} discussed a trade-off in moderation philosophies: nurturing versus punishing. Nurturing is described as 'an educational approach that aims at improving or reforming community members’ behavior (p.13)' \cite{jiang_trade-off-centered_2023}, while punishing focuses on 'ensuring that the person violating the rules receives consequences for their behavior (p.13)' \cite{jiang_trade-off-centered_2023}. Both philosophies have their merits in content moderation, but it is crucial for moderators to consider the context when determining the most suitable approach, or 'trade-off'. In our study, TikTok's moderation approach was predominantly punishment-oriented; it tended to emphasize the consequences of rule violations rather than the educational effects on the violators. However, this approach did not alleviate the conflict between BlindTokers and trolls; instead, it led to more severe outcomes, such as increased marginalization of BlindTokers.

It's important to note that many justice frameworks tend to classify stakeholders into categories of violators and victims. Utilizing this classification method, based on the judgments of the content moderation system, BlindTokers who used harsh language could be labeled as violators, and the trolls who instigated the conflict as victims. However, such a judgment fails to capture the full complexity of the situation. To gain a more nuanced understanding of this specific case and to propose viable solutions, we apply a transformative justice framework to analyze the experiences of BlindTokers and the harassment by trolls. Transformative justice is a holistic framework for analyzing justice \cite{salehi_sustained_2023, rabaan_survivor-centered_2023, sultana_shishushurokkha_2022, musgrave_experiences_2022}. It posits that harmful behaviors by offenders are often consequences of systemic injustice; rather than solely blaming and punishing individuals for violations and harm, transformative justice emphasizes identifying and addressing the systemic defects that contribute to such violations. In discussing the results within the context of transformative justice, we argue that a nurturing approach to moderation is more appropriate in the case of BlindTokers.

Transformative justice highlights two principles that are particularly pertinent to our study. The first is the identification of underlying factors at the societal or community level that contribute to violations. As detailed in the findings section, while the harsh language used by BlindTokers triggered content moderation, the actual root cause was the conflicts between the sighted audience and BlindTokers. These conflicts were, in turn, sparked by sighted people's misconceptions about BlindTokers. Digging deeper, it becomes evident that the lack of knowledge among sighted TikTokers about blindness was the initial trigger of these misconceptions. This step-by-step analysis reveals that the \textit{sighted people's lack of knowledge} about BlindTokers is a key underlying societal factor. Additionally, some BlindTokers reported that during these conflicts, certain sighted individuals made humiliating comments without facing moderation. This disparity may be attributed to differences in \textit{content moderation literacy}. We define content moderation literacy as the knowledge of how moderation systems operate, with a crucial aspect being the ability to communicate messages without triggering these systems. In our case, sighted trolls might have possessed higher content moderation literacy, enabling them to harass BlindTokers without triggering the system. Our findings suggest a possible explanation for this discrepancy. As discussed at the beginning of the discussion section, content moderation on TikTok is significantly influenced by visual aspects. The platform's inaccessible designs left most BlindTokers unaware of the moderation mechanisms, providing them with fewer opportunities to observe and reflect on content moderation compared to sighted individuals. This inaccessibility-induced disparity in content moderation literacy is another critical societal-level factor.

Another key principle of transformative justice involves proposing potential solutions that engage all stakeholders (in this scenario, BlindTokers, trolls, and the TikTok platform) to address societal-level problems. In this context, the TikTok platform represents 'society'. Therefore, our proposed solutions also focus on the socio-technical interplay among sighted TikTokers, BlindTokers, and the TikTok platform. Drawing inspiration from the nurturing moderation philosophy proposed by Jiang et al. \cite{jiang_trade-off-centered_2023}, we offer the following suggestions. First, we advocate for an educational approach aimed at enhancing the overall TikTok community's understanding of people with visual impairments. TikTok could organize events that promote participation from all community members, thereby raising awareness about the diversity of visual impairments. Second, we recommend adopting a more educational moderation strategy that alerts BlindTokers about the potential consequences of their posts before they publish harmful content, rather than imposing punitive measures post-publication. This approach could help BlindTokers expand their understanding of content moderation. Furthermore, we suggest that TikTok enhance its design to better capture and moderate subtle yet harmful content. This may require collaboration between BlindTokers and TikTok moderators. More comprehensive implications are discussed in the following subsection.

\subsection{Design Implications}
  

By integrating the transformative justice perspective with reactions to content moderation, we underscore the importance of several key implications. Firstly, there is a need to prevent triggers of moderation from both the public's actions and the platform's accessibility design. This involves not only addressing societal misconceptions and biases but also ensuring that the platform is designed with accessibility in mind to avoid inadvertently penalizing marginalized groups. Secondly, it is crucial to establish a balance between human and AI moderation. This balanced approach should be tailored to effectively meet the needs of diverse groups, especially in situations where content moderation has already occurred. Such a dual-focused strategy acknowledges the complexity of content moderation and aims to create a more inclusive and equitable online environment. We provide specific design implications as follows:

\subsubsection{Promote Moral Sensibilities of the Public towards Marginalized Groups}

Initiatives should be implemented to educate the public about disabilities, particularly visual impairments, to reduce societal misconceptions and biases. This could include awareness campaigns on TikTok, featuring content created by BlindTokers to offer insights into their experiences and challenges. Specifically, designers can adopt strategies to enhance the public's moral sensibilities \cite{garrett_felt_2023} by incorporating educational prompts alongside marginalized content. When users engage with content that touches on sensitive issues, integrated educational prompts can offer explanations about potential issues and consequences, thereby providing more context and fostering respectful discussions. This approach can prompt users to reconsider before engaging in harmful behaviors. For instance, TikTok could utilize the comment box feature effectively. Currently, the gray placeholder text in the comment box, which reads "add a comment...", could be altered to include reminders such as "be respectful, consider the video poster's identity and context." Such a subtle yet constant reminder in the comment box can serve as a cue for commentators to maintain decorum and thoughtfulness whenever they intend to comment on a video.


\subsubsection{Accessible Ways to Remind BlindTokers about Potential Violation}

Navigating the content creation process can be challenging, especially when inaccessibility leads to unintentional content violations. BlindTokers often remain unaware of having violated platform policies until after their content is removed. To alleviate the burden of content creation and mitigate potential negative consequences, designers can explore alternative mechanisms to alert BlindTokers about potential violations detected by the algorithm. Building on previous research, such as the use of vibration feedback in app design to accommodate visual impairments \cite{kamarushi_onebuttonpin_2022}, TikTok designers could implement similar features. For instance, they might introduce vibration notifications for users who self-identify as visually impaired or use hashtags related to BlindTokers. When the algorithm detects a potential issue, like a copyright violation or inappropriate exposure of certain body parts, the user could receive a vibration alert on their phone. Additionally, an audio notification could be employed, either reading out the specific concern or using a distinct ringtone, to inform users before posting that their content might violate platform policies (e.g., 'This video contains images that may violate copyright'). In addition, BlindTokers and other marginalized groups should have a say in how moderation policies are shaped and implemented. This could involve setting up advisory panels or forums where these groups can provide feedback and suggestions.

\subsubsection{Personalized Feeds with Accessible Resources For Better Knowledge of Content Moderation }

BlindTokers, as a marginalized group, encounter a wide range of comments on TikTok. Some of these are intentionally harassing, while others may not be. For interactions that seem attention-seeking or inflammatory, BlindTokers might choose to ignore or report them to avoid further conflict. In cases of misunderstanding regarding their identity, a more civil conversation could be initiated for clarification. As previously noted, there may be a disparity in content moderation literacy due to accessibility issues. To address this, the TikTok platform could offer more accessible resources that clearly present the rules, guidelines, and strategies to navigate content moderation. One practical approach could be to include tutorial videos in BlindTokers’ feeds, appearing periodically to assist them in avoiding conflicts and content moderation \cite{cai_after_2021}. These resources should particularly focus on helping BlindTokers recognize common types of trolls and provide techniques for avoiding conflicts with them.

\subsubsection{Balance Human-AI Collaborative Moderation to Meet Diverse Groups' Characteristics}

Humans and AI can collaborate more effectively in content moderation, especially in cases involving marginalized communities like BlindTokers. For instance, assigning straightforward tasks to AI while reserving contextually sensitive cases for human moderators represents an optimal approach to human-AI collaborative moderation \cite{lai_human-ai_2022}. This strategy underscores the importance of human intervention in handling issues related to marginalized groups. Moreover, our research has identified issues such as delayed or absent feedback on appeals and reports, or automated explanations that lack clarity. A recent study on AI moderation has proposed a human-user-AI collaboration model to better address users' needs and create more equitable algorithms \cite{schulenberg_towards_2023}. This model suggests that marginalized users, like BlindTokers, should have the opportunity to contribute input and influence AI design, ensuring that their specific needs are met. Thus, platforms and designers should consider incorporating human labor and actively collaborating with BlindTokers to enhance the effectiveness and fairness of their moderation algorithms and actions. Moderation systems should be equipped to consider the context in which content is posted. For instance, understanding the background of conflicts between BlindTokers and trolls can help in making more informed decisions regarding moderation. 


%% file: Sections/08_Limitations.tex
\section{Limitations and Future Work}

Despite our comprehensive efforts, we recognize that this study has limitations. A primary limitation is that our data consists solely of stories reported by participants, which may include subjective perceptions. Additionally, the data collection occurred between the summer of 2022 and the spring of 2023. Consequently, the findings might not accurately reflect the current state of TikTok’s moderation algorithms or its content moderation policies. However, we believe these experiences hold significant value for the Human-Computer Interaction (HCI) literature. They highlight an urgent societal issue within social media platforms and amplify the voices of BlindTokers, shedding light on their marginalized experiences. Such insights are crucial for understanding and addressing the challenges faced by visually impaired users in digital spaces.

We also propose directions for future research. As discussed in the related work section, studies investigating the experiences of blind users on short-video platforms have examined various cultural contexts, including Eastern (e.g., China \cite{rong_it_2022}) and Western cultures (as explored in this study). These studies collectively underscore the challenges blind users face in interactions with trolls. Notably, Chinese users tend to adopt a more passive approach to avoid conflicts, whereas Western participants in our study engaged more actively in such conflicts. These variations in social interaction styles point to different perceptions of identity among blind users across cultures. Therefore, future research should explore the specific impact of cultural factors on the experiences of blind users, or users with other disabilities, on short-video platforms. Understanding these cultural nuances can provide deeper insights into how identity and culture influence the digital experiences of users with disabilities.